\documentclass[a4paper,aps,prb,showpacs,twocolumn]{revtex4}
\usepackage{amsmath,amssymb,mathrsfs}
\usepackage[pdftex]{graphicx}

\begin{document}

\bibliographystyle{unsrt}

\title{Parameters of Tomonaga-Luttinger liquid in a quasi-1D material with Coulomb interactions}

\author{Piotr Chudzinski}
\affiliation{School of Mathematics and Physics, Queen's University of Belfast}

\date{20.02.2020}

\begin{abstract}

In this work we derive a new scheme to calculate Tomonaga-Luttinger liquid (TLL) parameters and holon (charge modes) velocities in a quasi-1D material that consists of two-leg ladders coupled through Coulomb interactions. Firstly, we obtain an analytic formula for electron-electron interaction potential along the conducting axis for a generalized charge distribution in a plane perpendicular to it. In the second step we introduce many-body screening that is present in a quasi-1D material. To this end we propose a new approximation for the charge susceptibility. Based on this we are able to find the TLL's parameters and velocities. We then show how to use these to validate the experimental ARPES data measured recently in p-polarization in $NbSe_3$. Although we focus our study on this specific material it is applicable for any quasi-1D system that consists of two-leg ladders as its basic units.

\end{abstract}

\maketitle


Probing spin-charge separation in quasi-1D materials has been one of the long-standing challenges in experimental solid state physics of strongly correlated systems. More than two decades\citep{CLAESSEN1995121} of attempts has produced more false alarms than proper findings and, as a result, most recent experimental evidences are combined with a theoretical estimate of e.g. TLL collective modes' velocities\citep{Ma_2017},\citep{osti_889174},\citep{BOUNOUA201849}. A strong emphasis is then put on the fact that the experiment and theory are in a reasonable quantitative agreement. While such procedure has been outlined for a Hubbard-like models\citep{Ma_2017},\citep{osti_889174},\citep{BOUNOUA201849}, with short range interactions, there is no such a scheme available for other class of realizations, actually more frequent, i.e. systems with long range interactions. This work aims to fill this gap, and to do it by providing analytical expressions for the desired quantities, which, contrary to numerical results, are easily transferable from one case to another. So far, among these class of models, only carbon nanotubes were investigated in detail, see seminal work Ref.\cite{GogolinEgger}, here we extend this work to apply it to systems with lower local symmetry. 

The successful search for the TLL is facing two conditions \textbf{i)} and \textbf{ii)}. The frequently used model with a short range Hubbard U-V interactions is valid for a quasi-1D material where the other 2D (or 3D) bands, that are not crossing Fermi energy $E_F$ but provide screening, fall very close enough (a fraction of a bandwidth) to the $E_F$. However, the opposite situation is more frequent, where the other bands overlap (in energy) only with the very bottom of the 1D band \textbf{i)}. This gives a broader energy window of a weak screening. Secondly, \textbf{ii)}, a narrow wire (on atomic scale) that hosts only a single 1D channel is very prone to undergo Peierls instability (if sufficiently close to any incommensurability) or Wigner instability (if carriers in a given sub-band are sufficiently dilute) both destroy the TLL phase. This tendency can be minimized for a larger conduction band electrons' orbital, spread over a few atoms in a unit cell, but then a local symmetry of the orbital will be lowered. The two leg ladder will then be the simplest, and actually the most common, case. This is the situation that we aim to tackle in this paper. 

Our study is motivated by a recent experimental report, Ref.\cite{PhysRevB.99.075118},\cite{valbuena2016emerging}, of 1D TLL states observed in p-pol in NbSe$_3$. We can summarize those findings as: i) at each Fermi point there are \emph{two} linear dispersions\cite{valbuena2016emerging}, at the inner band (following Ref.\cite{PhysRevB.99.075118} we shall call this anti-bonding band A1 hence the Fermi point $k_{F1}$) the two velocities converge to the same value, at outer band A3 the two velocities are different (one $A3_s$ is equal to those at $k_{F1}$ while the other $A3_h$ is a factor $\approx 1.25$ larger)  ii) the spectral function can be fitted with a finite temperature expression\cite{STM-finiteT} for the TLL $A(\omega,T)$ with the characteristic Green's function exponent $\alpha=0.24$\cite{valbuena2016emerging}. It is surprising that with such a small ratio of velocities there is such a substantial value of $\alpha$ exponent and our aim here is to explain this experimental finding.        

The manuscript is organised as follows. We start by deriving Coulomb interactions potential along a low symmetry 1D column. With this we compute the TLL parameters for a single column and for the entire quasi-1D material with a partial screening incorporated in the solution. Then we focus on a concrete realisation and analyse the result of the ARPES experiment\cite{PhysRevB.99.075118},\cite{valbuena2016emerging}.  

\paragraph{Tomonaga-Luttinger liquid} The Hamiltonian of the TLL state is written in terms of fluctuations of these collective modes:
\begin{equation}\label{eq:ham-TLL-def}
    H^{1D}[\nu]= \sum_{\nu} \int \frac{dx}{2\pi}
    \left[(v_{\nu}K_{\nu})(\pi \Pi_{\nu})^{2}+\left(\frac{v_{\nu}}{K_{\nu}}\right)(\partial_{x} \phi_{\nu})^{2}\right]
\end{equation}

where $\nabla\phi_{\nu}(x)$ gives the local density of fluctuation $v_{\nu},K_{\nu}$ are velocity and TLL parameter ($\sim$compressibility) of a given bosonic mode $\nu$, these
depend on electron-electron interactions with small momentum
exchange. In the simplest approximation $K_{\nu}\approx (1-2g)$ where $g$ is a strength of the  electron-electron Coulomb interaction $q\rightarrow 0$ and assuming that Galilean invariance holds $v_{\nu}K_{\nu}=V_F$. The $V_F$ is a Fermi velocity $\sim 2t_{b}$ (where we take tight binding model with 1D chains arranged along the b-axis) which also which determines the UV cut-off of our theory $ 2  t_b\sim \Lambda$. One immediately realizes that pure Coulomb interactions diverges when $q\rightarrow 0$ which makes $g$ an ill-defined quantity while in a multimode case (each with different $v_{\nu}$) the Galilean invariance in not guaranteed. In this paper we aim to improve this simple, approximate formula for $v_{\nu}$.   

When the crystal's columns, on which the gapless states exist, are grouped into pairs and so there are two bands crossing Fermi energy then a two leg ladder description applies. There are four bosonic modes $\rho\pm$, $\sigma\pm$ as both spin and charge can oscillate symmetrically or anti-symmetrically within the two legs of the ladder, we call it later total and transverse modes respectively. 
In the most general case one expects the following spectrum: four linear dispersions that start at each Fermi point. If we assume that there is no high-energy phase transition induced by changing range of the interaction then the two charge modes shall cross at the $\Gamma$ points (forming structures that each resembles a Dirac point with constant velocities) while the two spin modes shall have parabolic behaviour (hence $v_{\sigma\pm}(q_{||}\rightarrow \Gamma)\rightarrow 0$) close to the $\Gamma$ point \footnote{Strictly speaking this is what an exact analytic solution for Hubbard model tells us when hopping between chains is not extremely strong (on-chain physics dominates on-rung physics) but the numerical results for extended Hubbard models suggest that there is no qualitative modification of this spectrum as the range of interaction is changed}.   

\begin{figure}
		\centering
		\includegraphics[width=1.1\columnwidth]{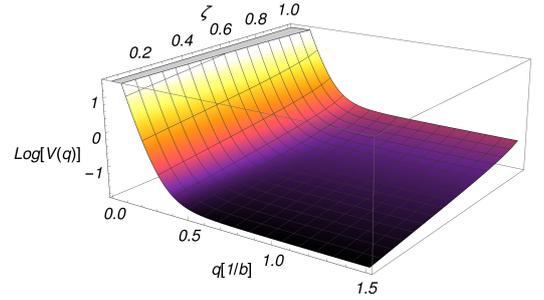}\\
	\caption{Fourier transform of the Coulomb interaction Eq.\ref{eq:inter-eff-real} shown as a function of momentum and parameter $\zeta$ that describes inhomogeneity (eccentricity) of the $\rho(\vec{r})$ cross-section.}\label{fig:CoulombType}
\end{figure}

\paragraph{Single ladder} The first step, in quantitative treatment of Coulomb interactions, is to find their dependence as a function of inter-carriers' distance. We build our reasoning on the top of some single-particle DFT calculations, which presumably provides us with electronic density of the carriers occupying the 1D band. What such calculations are unable to capture are electron-electron correlations inevitably present in 1D many-body state described by Eq.\ref{eq:ham-TLL-def}. We seek a way to determine $v_{\nu},K_{\nu}$ from the knowledge of single-particle physics. 
The bare Coulomb potential for a 1D column of charge density reads\cite{GogolinEgger}
\begin{equation}\label{eq:Coulomb1}
\bar{V}_{\rm Coul}(\vec{r}-\vec{r'}) = \frac{e^2/\kappa}{\sqrt{(x-x')^2+4R^2\sin^2((y-y')/2R)+d^2}}
\end{equation}
where the $d$ is a thickness of the charge layer. The scattering amplitude (that enters to second quantization Hamiltonian) of the Hartree type interaction is given by an integral over the elementary unit cell: 
\begin{equation}\label{eq:2nd-quant-ampl}
V(\vec{r},\vec{r'})= \int d\vec{r} \Psi^*(\vec{r})\Psi^*(\vec{r'})\bar{V}_{\rm Coul}(\vec{r}-\vec{r'})\Psi(\vec{r})\Psi(\vec{r'})
\end{equation}
where $\Psi^*(\vec{r}),\Psi^*(\vec{r'})$ are wave-functions of interacting electrons, a homogenous Bloch waves along the x-direction. We assume that in a perpendicular plane the charge density is spread over a section of a distorted toroid. We add an extra parameter $\zeta$ that accounts for an inhomogeneity along the circumference of the toroid where $\zeta=0$ correspond to the symmetric homogeneous distribution (and a constant radius) like in nanotube. Hence we generalize expression given in Ref.\cite{GogolinEgger} with symmetry reduced down to $C_4(\vec{b}),S_4(\vec{b})$. We integrate over perpendicular coordinates to get an interaction amplitude along the b-axis $V(x)$:
\begin{equation}\label{eq:toroid}
V(x)=\int_{\phi R}^{2\pi R}\int_{\phi R}^{2\pi R} \frac{dy}{2\pi R} \frac{dy'}{2\pi R} \frac{\bar{V}_{\rm Coul}(\vec{r}-\vec{r'})}{1-\zeta\sin((y-y')/2R)}
\end{equation}
The closed analytic form for the integral is known also in this more general case:
\begin{equation}\label{eq:inter-eff-real}
V(x)=\bar{U}\frac{\sqrt{1-\zeta^2} \Pi \left(\phi ;\zeta \left|\left(\frac{2 R}{\sqrt{d^2+4 R^2+(x-x')^2}}\right)^2\right.\right)}{\left(\phi \sqrt{d^2+4 R^2+(x-x')^2}\right)}
\end{equation}
where $\Pi(\phi;\zeta|1/\tilde{x})$ is the incomplete elliptic
integral of the third kind. The integral is parametrized by $\bar{U}=U/N$ (chosen appropriately depending on the ab-initio method) e.g. for cRPA we account for all screening provided by carriers residing on all other orbitals but not the 1D d-orbital which is equivalent of saying that at a given UV cut-off $\Lambda$ (usually proportional to the inverse lattice spacing) the correlation interactions saturate to a value given by local Hubbard-type repulsion. The parameters $\phi$ (angle of the sector of the toroid) and $\zeta$ (distortion of the toroid) are determined by the
geometry of the given \emph{eigen-orbital} in the a-c plane perpendicular to 1D axis. The prefactor ensures that no matter what the geometry is, there total density of charge is normalized.   
We need to perform Fourier transform of $V(x)$ which will complete our description of the Hartree term in a single ladder/column, the result is illustrated in Fig.\ref{fig:CoulombType}. 

For Coulomb electron-electron interactions there shall be also exchange interactions. 
Their value, for $r=a$ is again known from the \emph{ab-initio} calculations and parametrized by $J(r=a)=J$. This can be thought of as an interaction of a carrier with an exchange hole of another carrier. The exchange hole is described by a two-body correlation function $g(x,x')$. This extra term will convolve one of the wave-functions in Eq.\ref{eq:2nd-quant-ampl}. The two-body density, in a transnationally invariant TLL, has a dominant contribution that scales like $(r-r')^{-(1/4(1/K_{\rho+}+3/K_s)+2)}|_{K_s=1}$.
For a sufficiently strong repulsive interactions $K_{\rho+}\ll 1$, we obtain a quantity that rapidly decays in real space which agrees well with our intuitive understanding of exchange. Then the overall Fourier transform $J(q)$ will be the previous Fourier transform of screened Coulomb potential times an increasing power law and so the $J(q=2k_F)$ may be substantial (please note that this enters only to $g_2$ and does not enter to backscattering terms $g_1$).   

\paragraph{Quasi-1D material} After we have established the parameters of a single ladder, we shall now move on to a quasi-1D material -- a system that consist of multitude of parallel ladders.
 The full modelling has to account for the fact that we are dealing with a set of 1D systems coupled by long range density-density (forward scattering) interactions, these inter-ladder (charge) density-density terms are non-negligible, they can be as large as $t_b/2$\cite{PC-bronze}. We first compute screening of a single TLL by other columns which cuts the divergence of $V(q)$ when $q \rightarrow 0$. To this end we take $V_{eff}(q)=V(q)/(1+G(q,T)\chi_{TLL}(q))$. Here we take an RPA approximation for the dielectric function. For the charge susceptibility of the two leg ladder with one of the velocities $v_{\rho+}$ much different than $V_F$ and all other velocities $v_{\rho-,\sigma\pm}\approx V_F$ (see below) and $K_{\rho-,\sigma\pm}\approx 1$ we can generalize the result obtained in Ref.\cite{IucciFieteGiam}
\begin{widetext}
\begin{multline}\label{eq:2leglad-susc}
\chi_{TLL}(q)= 4\pi \frac{\Gamma\left(1-\left(\frac{K_{\rho+}}{4}+\frac{3}{8}\right)\right)}{\Gamma\left(\frac{K_{\rho+}}{4}+\frac{3}{8}\right)}\frac{|\omega^2-V_F^2q^2|^{\left(\frac{K_{\rho+}}{4}+\frac{3}{8}\right)-1}}{V_F^{\left(\frac{K_{\rho+}}{2}+\frac{3}{4}\right)-1}}\exp\left(-\imath\pi[\left(\frac{K_{\rho+}}{4}+\frac{3}{8}\right)-1]\mathfrak{H}_{\Theta}[\omega^2-V_F^2q^2]\right)\\
 F_1\left(\frac{K_{\rho+}}{4},\frac{K_{\rho+}}{4}+\frac{3}{8}-\frac{1}{2},1-\left(\frac{K_{\rho+}}{4}+\frac{3}{8}\right),\frac{K_{\rho+}}{4}+\frac{3}{8};1-\left(\frac{v_{\rho+}}{V_F}\right)^2,1-\frac{\omega^2-v_{\rho+}^2q^2}{\omega^2-v_{F}^2q^2}\right)|_{\omega\rightarrow\Lambda^{IR}}
\end{multline} 
\end{widetext}
where $F_1()$ is the Appell hypergeometric function and $\mathfrak{H}_{\Theta}$ is a Heaviside theta function that appears upon analytic continuation. Since we  are interested in the smallest $q$ and static response we work within the radius of convergence of the Appell hypergeometric function. For the $G(q,T)$, which accounts for a local field corrections we take $G(q,T)=\bar{G}(q,T)/(q^2 + \Lambda_{IR}^2)$ which is a screened Coulomb interaction times the static structure factor, the lesser correlation function of the screening medium (it is by definition what local field correction is supposed to capture). We take $\bar{G}(q,T) \sim A(q,\omega\rightarrow 0,T;\Delta\equiv \Lambda_{IR})$ where the spectral function is known from Ref.\cite{EsslerTsvelik} and the IR cut-off $\Lambda_{IR}=max(T,t_{\perp})$. Please note that with this choice the entire $G(q,T)$ can be interpreted as a propagator (i.e. a retarded correlation function in the Lehmann representation) of the bosons (the holons) carrying interactions in between the 1D wires. Indeed a sole 1D ladder cannot self-screen itself (and RPA expression is exact therein, local field corrections are absent) so entire local field correction must originate from its environment. This approach, based on the $t_{\perp}$ tunnelling, is sufficient to limit the increase of $V_{eff}(q\rightarrow 0)$ to a well defined constant value, which among other implication also gives a finite, non-divergent holon's velocity. 

\begin{figure}[h]
	\centering
	\includegraphics[width=1.\columnwidth]{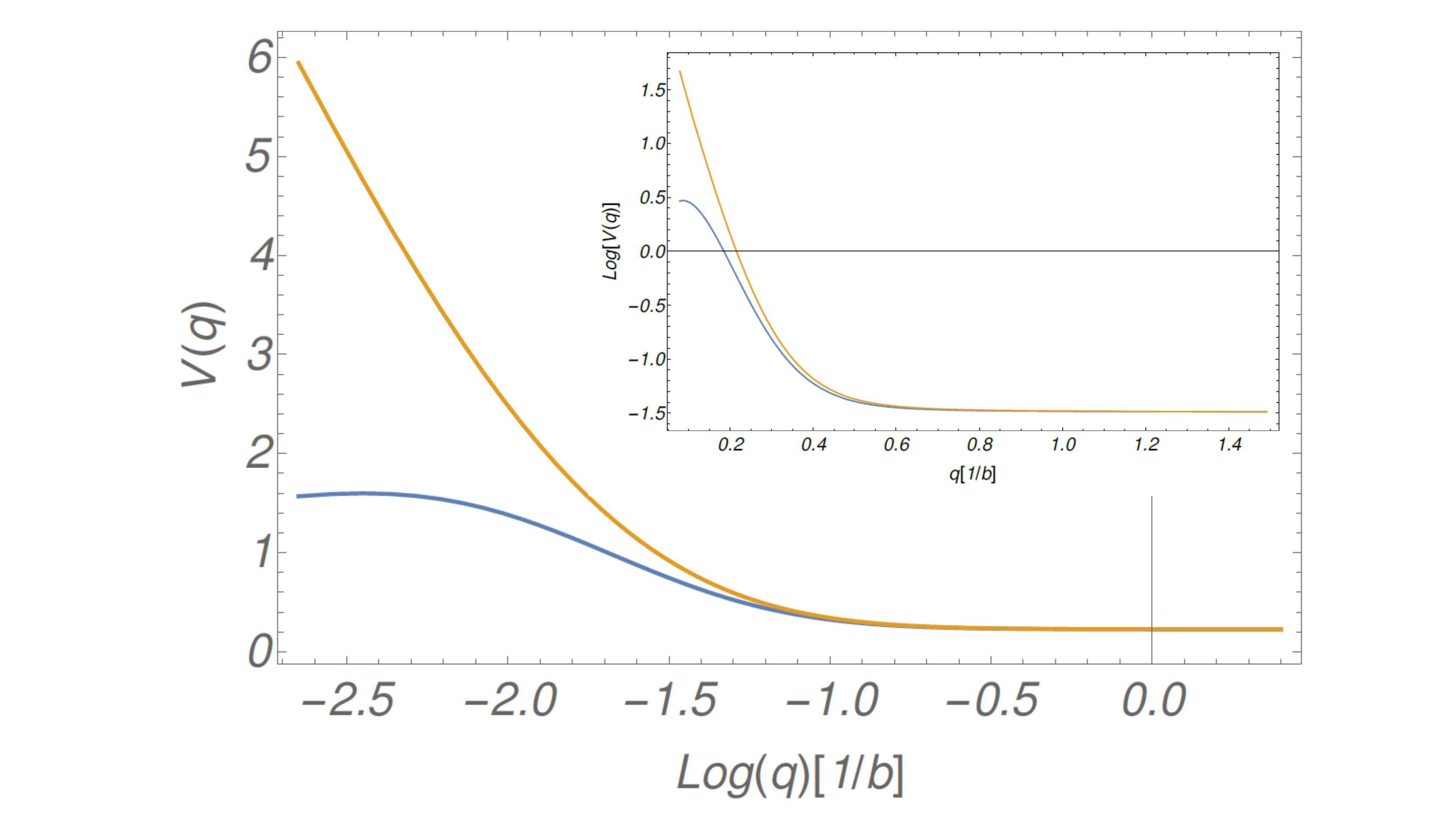}\\
	\caption{Screening effects in quasi-1D material. Logarithm of electron-electron interactions as a function of momentum q is shown. Yellow curve is for bare non-screened $V_{Coul}(q)$, a Fourier transform of Eq.\ref{eq:inter-eff-real}, while blue line is screened by the susceptibility given in Eq.\ref{eq:2leglad-susc}. The main plot is in log-linear coordinates and manifest the asymptotic$\sim Log(q)$  behaviour while the inset is in linear-log coordinates to show the real momentum dependence.}
	\label{fig:screened1D}
\end{figure}  

\paragraph{TLL parameters $K_{\nu}$} The bosonization is made in $0/\pi$ bands so one begins with Hartree and Fock intra and inter-band interactions. In each band we have Hartree $V(q)$ and Fock $J(q)$ and then it is known\cite{Gogolin99bosonizationand} that:
\begin{multline}
g_{4mn}= 1/2(2V^{(mn)}(q=0)-J^{(mn)}(q=0))/\sqrt{V_{Fm}V_{Fn}}\\
g_{2mn}= ([V^{(mn)}(q=0)+J^{(mn)}(q=2k_{Fo/\pi})]\\
 -1/2[V^{(mn)}(q=2k_{Fo/\pi})+J^{(mn)}(q=0)])/\sqrt{V_{Fm}V_{Fn}}\\
g_{\sigma mn}= -J^{(mn)}(q=0)/\sqrt{V_{Fm}V_{Fn}}
\end{multline} 

where $m,n=o,\pi$ and we took into account a possibility of different Fermi velocities in both bands. The band dependence of interaction comes from their different shapes\footnote{on which we elaborate in Supp.Mat for the specific case of NbSe$_3$}. Once interactions are determined in a band-basis one performs $S[\pi/4]$ rotation\citep{PC-basis-RG} to the total-transverse basis $g_{\nu}$. The TLL parameters and velocities for the neutral modes follow directly from here and can be given by the following, single ladder, formula: $K_{\nu}=\sqrt{\frac{1-g_{\nu}}{1+g_{\nu}}}$ where $g_{\rho-}=(g_{2o}-g_{2\pi})+J_{oo}(2k_F)$, $g_{\sigma+}=g_{\sigma o}+ g_{\sigma\pi}$ and $g_{\sigma-}=g_{\sigma o}-g_{\sigma \pi}$. In the specific NbSe$_3$ case all these three $K_{\nu}$ are close to $1$ and so $v_{\nu}\approx V_F$. In the bosonic language when interactions are spin/orbital independent at $q\rightarrow 0$ only the $\nabla\phi_{\rho+}$ operator is coupled with the long range interactions. Since $g_{oo}\approx g_{\pi\pi}\approx g_{o\pi}$ the $g_{\rho+}$ is by far the largest, but since it is further modified by inter-chain interactions we discuss it separately below.

\begin{figure}[h]
	\centering
	\includegraphics[width=1.\columnwidth]{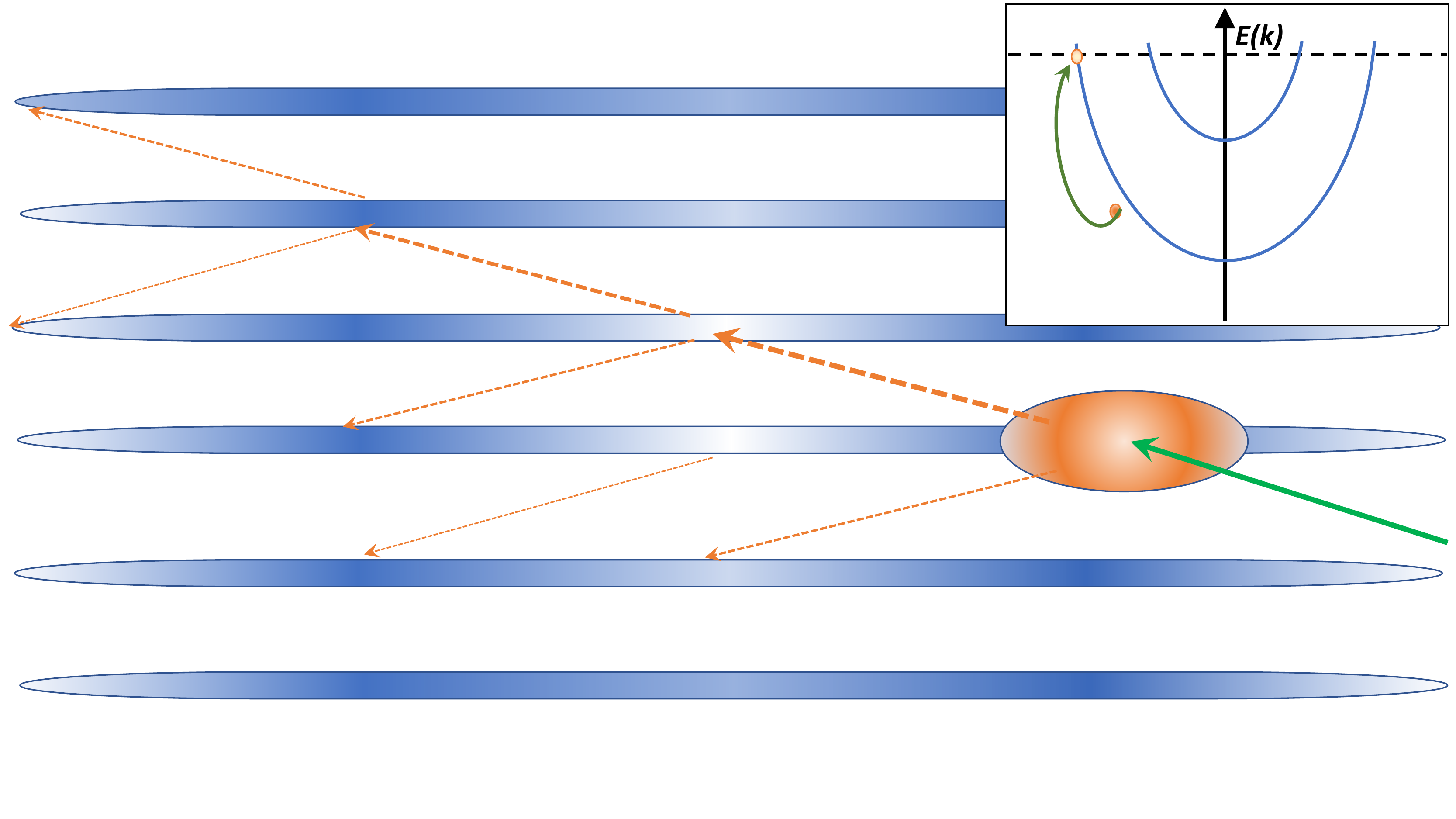}\\
	\caption{An ARPES photo-emission event shown in the real space. We see that a photon (green arrow) incoming with a finite $q_{\perp}$ shall induce charge density fluctuations not only in the 1D ladder where photo-hole (orange oval) has been instantaneously created but (through screened Coulomb interactions, orange arrows) in the surrounding 1D systems. An inset shows the same event in a reciprocal space: instantaneous recombination of a photo-hole with a fermion at one of the Fermi points (together we call this object a photo-density).}
	\label{fig:ARPESmodel}
\end{figure}

\paragraph{Holon mode: Connection with experiment} Now we account for a finite aperture size of the ARPES experiment, which implies that the final state is a mixture of waves with different $q_\perp$  (this is a coherent state that results from Fresnel diffraction of electronic waves\cite{Chudzinski_2019}, see Fig.\ref{fig:ARPESmodel}) The problem is simplified because again only $\rho+$ modes are exposed to these long-rage interactions. The treatment of this system can then be based directly on the solution provided by H.J.Schulz in Ref.\citep{Schulz_1983}. 
In that work it was found that the velocity is equal to:
\begin{equation}\label{eq:vrho-multicol}
v_{\rho+}= \frac{1}{\Delta q_{\perp}}\int_0^{\Delta q_{\perp}} dq_{\perp} V_F \sqrt{(1+g_{\rho+}^{(tot)})(1+2g_{4\rho+}-2g_{2\rho+})}
\end{equation}
where $\tilde{g}_{\perp}=\frac{g_\perp}{\epsilon_{||}^{(0)}(1+\epsilon^{(0)}_{\perp}(q_{||}/q_{\perp})^2}$, $g_{\perp}=V_{\perp}/V_F$, $g_{4\rho+}=\sum g_{4mn}$, $g_{\rho+}^{tot}=2g_{4\rho+}+2g_{2\rho+}+\tilde{g}_{\perp}$\footnote{here we profit from our treatment of $V_{eff}$ and based on result in Fig.\ref{fig:screened1D} we take $\epsilon_{||,\perp}(q_{||})\sim (q_{||}^2+m^2)^{-1}$ which is equivalent to a model with an infinitesimally small many body gap $m$}). We take an integral over a finite $\Delta q_{\perp}$ in order to account for the finite aperture of the (Fresnel zone focused) nanoARPES device in Ref.\cite{PhysRevB.99.075118},  here we take $\Delta q_{\perp}=\pi/6$\footnote{M.A.Valbuena private communication}. At the same time the exponent of the single particle Green's function (momentum integrated) is equal to $\alpha=(C_{\rho+}+C_{\rho+}^{-1})/8+6/8-1$ where $6/8$ comes from the three bosonic modes with $K_{\nu}\approx 1$ and:
\begin{equation}\label{eq:Crho-multicol}
C_{\rho+}=\sqrt{\frac{1+2g_{4\rho+}-2g_{2\rho+}}{1+2g_{4\rho+}+2g_{2\rho+}}}\left(1+\frac{\pi}{8}y_{\perp}ln[\frac{y_{\perp}}{4}]\right)
\end{equation}
where $y_{\perp}=\frac{g_{\perp}}{1+2g_{4\rho+}+2g_{2\rho+}}$The above formulas, with a physically sensible $g_{\perp}\in (0.4,0.6)$ gives the following estimates for observable quantities $v_{\rho+}\in (1.2,1.35)V_F$ and $\alpha\in (0.15,0.2)$. 

\paragraph{Relative intensity of the four modes} The formula Eq.\ref{eq:inter-eff-real} allows not only to compute parameters of the stationary TLL, but also to study the dynamics of an excited photo-hole. This in turn gives an insight into relative intensities measured for various TLL modes. To investigate the relative intensities of the four bosonic dispersions we take the three step model for photo-emission. Contrary to the Fermi liquid, in TLL for each momenta we have several (four in our case) bosonic branches and the strength of each relaxation channel is proportional to an overlap integral between a photo-density $\rho_i(\vec{r})$ (defined below) and a given collective excitation of TLL. Initially, just after the photo-emission event, the photo-hole (a single fermion state) is created in an intermediate state $\psi_{i}(\vec{r})$ that is not an eigenstate but a localized combination of many excited states of the system. Then this initial excitation relaxes into collective, bosonic eigen-states of the many-body system. 
The relaxation is slower than for the single-particle bands and happens through Coulomb interaction \cite{PhysRevLett.105.226407}, the one given in Eq.\ref{eq:inter-eff-real}, i.e.
\begin{equation}\label{eq:relax}
H_{rel}=\sum_{k_{phot},q} \int dq d\omega T(1,2)\rho(q,\omega)c_{bi}(k_F)c_i^{\dag}(k_{phot})
\end{equation}
where we anihilate the photo-hole by re-combination with an electron from the Fermi surface. The energy-momentum released in this decay process goes into a particle-hole collective density excitation, the eigen-state of the TLL, through electron-electron interaction $T_{1,2}\sim V_{eff}(x)$. The last two anihilation-creation operators in Eq.\ref{eq:relax} can be together called the \emph{photo-density} $\rho_i(\vec{r})$ $\equiv c(k_F)c_i^{\dag}(k_{phot})$. The $\psi_{i}(\vec{r})$ is located on atomic orbitals that fulfil dipole matrix ARPES selection rules and among these has to have a spatial distribution matching the evanescent wave of the emitted photo-electron, to ensure their large overlap. The photo-electron wave-function can be computed by inverse scattering method\cite{Chudzinski_2019} and from this calculation we know that $\psi_{i}(r_\perp)\approx H_0(r_\perp)$ inside the 1D wire (that acts as a trap). Since the Hermite polynomial $H_0(r_\perp)$ is a symmetric Gaussian then the overall symmetry of $\rho_i(r_\perp)$ is determined by a wave-function of the $c_{A1,3}(k_F)$ i.e. $\psi_{A1,3}(\vec{r};k_{F1,3})$ is a wavefunction (at the Fermi level) of a matching band     

The amplitude of the process, when resolved among different bosonic modes, reads:
\begin{equation}
 T(1=i,2=\nu)= \int d r_{\perp} V_{eff}(\vec{r})\rho_{\nu}(\vec{r})\psi_{i}(\vec{r})\psi_{A1,3}(\vec{r};k_{F1,3})
\end{equation} 
 
where the last two terms taken together give $\rho_i(\vec{r})$ and  $\rho_{\nu}(\vec{r})=\nabla\phi_{\nu}$. Different TLL modes have different symmetry with respect to the middle horizontal plane of the two leg ladder: the total modes $\nabla\phi_{\rho+}, \nabla\phi_{\sigma+}$ describe oscillations in phase of the densities of the two legs of the ladder while the transverse modes $\nabla\phi_{\rho-}, \nabla\phi_{\sigma-}$ describe oscillations in anti-phase. On the other hand, the difference between the two DFT bands is that one is symmetric (the bonding one A3 in $NbSe_3$) and the other anti-symmetric (anti-bonding one A1 in $NbSe_3$) on the rung that links the two legs of the ladder. Hence we can deduce that $\rho_i(\vec{r})$: for $(\omega,k)$ emission point close to the A3 (bonding band) dispersion $\rho_i(\vec{r})$ is an electron-hole described as a wave with two legs oscillating in phase, while for $(\omega,k)$ close to the A1 (anti-bonding) band $\rho_i(\vec{r})$ is a hole's wavefunction with an extra $\pi$ phase (or a '-' sign) between the two legs of the ladder. Since potential given by Eq.5 is symmetric and does not allow to mix $(\omega,k)$ points then in the first case relaxation into TLL eigenstates of the total holon/spinon modes is much more probable, while in the second case relaxation into transverse holon/spinon modes is much more probable. Let us emphasize that this selection rule is different from the standard dipole matrix selection rule frequently used in polarization dependent ARPES, also used in NbSe$_3$. The latter rule relates polarization vector with final states in the sample \emph{and in the detector}, while in the present case the selection is between several emergent collective solutions of a many-body problem (several eigen-states at the same $\vec{q}$). Please note that this selection rule is based on the fact that $V_{eff}(q)$ is strongly decaying with increasing $q$. 

Overall we predict that at each Fermi point ARPES will detect \emph{two} (instead of four) linear dispersions starting at $k_{F3}$ and $k_{F1}$.  The two branches are distinguishable when the dispersion approaches the $\Gamma$ point of Brillouin zone, since one is linear (charge) and one is parabolic (spin). The velocity of the spinon branch at $k_{F3}$ (total spin mode) is expected to be equal to velocities of the transverse modes visible at $k_{F1}$. The holon branch visible at $k_{F3}$ should have velocity approximately 25$\%$ higher than other dispersions. 

\paragraph{Discussion} Situation is even more complex if the TLL holon dispersion crosses a single-particle band as it happens e.g. with band C2 in NbSe$_3$\citep{valbuena2016emerging}. Then the photo-hole may relax much faster into these single particle states especially in the vicinity of the crossing point. This shall manifest in the experiment since at the crossing with the dispersion of the single electron b2 band we expect that $\psi_{i}(\vec{r})$ will sink out into this auxiliary dispersion so the intensity of the bosonic branches will be diminished. When energy of the photo-hole is larger that that of electron's bound within the band C2, then a two stage process is possible: in the first stage a single particle from the C2 falls onto a photo hole state and later the b2 hole relaxes onto Fermi level producing the final bosonic excitation $1_i \rightarrow 1_{b2+\nu} \rightarrow 2_{\nu}$. This is a higher order process with two \emph{intermediate e-h} involved, but it may be favoured by the faster C2 relaxation rate. Then the selection rule described above does not apply any more, instead the amplitude is proportional to $T(1_i , 1_{b2+\nu}, 2_{\nu})$. Experimentally this recovery of $\rho+$ mode intensity in the anti-symmetric band (e.g. A1 in NbSe$_3$) will manifest as an emergence of a second, shifted Dirac cone ($\approx 0.1eV$ below the bottom of the $\rho-$ cone) with a velocity $v_{\rho+}$ hence substantially larger than $V_F$. This is precisely what has been observed in the experiment, Ref.\cite{PhysRevB.99.075118},\cite{valbuena2016emerging}.  

The results presented so far are for a ladder coupled through single-particle hybridization and inter-ladder Coulomb couplings. However our method can be also applied to an interaction coupled ladder where each constituent has the symmetry lowered down to $C_4(\vec{b}),S_4(\vec{b})$ so the entire ladder can have only $C_2(\vec{b}),S_2(\vec{b})$ symmetry. However, in the interactions $V_{\perp}^{(intra)}$ coupled ladders both $K_{\rho\pm}$ are determined through a split $\sim V_{\perp}^{(intra)}$ of the single leg $K_{\rho}$ so we shall have two modes that are far from $K_{\nu}=1$ and so two modes that enter Eq.\ref{eq:vrho-multicol}-Eq.\ref{eq:Crho-multicol}. With small modifications, result of current work can be also applied to systems that are recently under intense scrutiny: artificial 1D systems created (or self-organized) on a dielectric surface. On a surface the interactions are substantially less screened, hence the necessity to tackle the system of electrostatically coupled wires with strong $g_{\perp}$ and weak $t_{\perp}$. 
Here it has to be noted that our approach is complementary to the in Ref.\cite{PhysRevB.79.235321} where they have modelled a single wire of a substantial perpendicular size (a mesoscopic charge density with relatively long decay length in the perpendicular direction). However attempts to bring such a mesoscopic model to sub-nm domain, while neglecting entire atomic/orbital physics, has to be excluded as unreliable and our microscopic approach offers a much better approximation in this case. 

In conclusion we have given a closed analytic expression for density-density interactions in quasi-1D materials. With input parameters known from DFT calculations, such as conduction orbital shape and strength of local interaction (effective Hubbard U e.g. from cRPA), one is able to give the value of the TLL parameter for the holon modes. We also analyse in detail the outcome of the recent ARPES experiment. For concreteness, we apply our results to a recently studied material NbSe3, however our reasoning presented here, can serve as a benchmark for future ARPES studies of similar quasi-1D compounds.

\paragraph{Acknowledgements} I would like to sincerely thank Thierry Giamarchi, Marco Grioni, Stephane Pons and Miguel A. Valbuena for numerous discussions that inspired this study and, in particular, Enric Canadell for sharing his vast knowledge on quasi-1D materials.   


\bibliography{NbSe2nd}

\end{document}